\documentclass[aps,preprint,amsmath,amssymb,11pt]{revtex4}
\usepackage{graphicx}
\usepackage{epstopdf}
\usepackage[T1]{fontenc}
\begin{document}

\title{Search for Top Quark FCNC Couplings at Future Circular Hadron Electron Collider}
\author{H. Denizli}
\email[]{denizli_h@ibu.edu.tr}
\author{A. Senol}
\email[]{senol_a@ibu.edu.tr} 
\affiliation{Department of Physics, Abant Izzet Baysal University, 14280, Bolu, Turkey}
\author{A. Yilmaz}
\email[]{aliyilmaz@giresun.edu.tr}
\affiliation{Department of Electric and Electronics Engineering, Giresun University,  28200, Giresun, Turkey}
\author{I. Turk Cakir}
\email[]{ilkay.turk.cakir@cern.ch}
\author{H. Karadeniz}
\email[]{hande.karadeniz@giresun.edu.tr}
\affiliation{Department of Energy Systems Engineering, Giresun University, 28200, Giresun, Turkey}
\author{O. Cakir}
\email[]{ocakir@science.ankara.edu.tr}
\affiliation{Department of Physics, Ankara University,  06100, Ankara, Turkey}
\begin{abstract}
A study of single top quark production via flavor changing neutral current interactions at $tq\gamma$ vertices is performed at future circular hadron electron collider. The signal cross sections for the processes $e^{-}p\to e^{-}W^{\pm}q+X$ and $e^{-}p\to e^{-}W^{\pm}bq+X$ in the collision of electron beam with energy $E_e=$ 60 GeV and proton beam with energy $E_p=$ 50 TeV are calculated. In the analysis, the invariant mass distributions of three jets reconstructing top quark mass, requiring one b-tagged jet and other two jets reconstructing the $W$ mass are used to count signal and background events after all selection cuts. The upper limits on the anomalous flavor changing neutral current $tq\gamma$ couplings are found to be $\lambda_q < $ 0.01 at future circular hadron electron collider for $L_{int}=100$ fb$^{-1}$ with the fast simulation of detector effects. Signal significance depending on the couplings $\lambda_q$ is analyzed and an enhanced sensitivity is found to the branching ratio BR($t\to q\gamma$) at the future circular hadron electron collider when compared to the current experimental results.
\end{abstract}

\maketitle

\section{Introduction}
One of the characteristic features of top quark which makes it very interesting is its large mass. 
Precise measurements of the couplings among top quark, gauge bosons and quarks are sensitive test of new physics (search for deviations) Beyond the Standard Model (BSM). The cross section for single top quark production via electroweak interactions is about three times smaller than the pair production which can be produced by strong interaction process at the Large Hadron Collider (LHC). Top quark interacts primarily by the strong interaction, but only decays through the weak interaction to a $W$ boson and a bottom quark (most frequently). It provides unique probe to search for the dynamics of electroweak symmetry breaking. With the high rates, it has the potential for precision studies.

 The Flavor Changing Neutral Current (FCNC) transitions are not present at the lowest order and suppressed at loop level due to the GIM mechanism in the Standard Model (SM) \cite{Glashow70}. Therefore, the top quark FCNC interactions would be a good test of new physics at the present and future colliders. In the BSM scenarios such as two-Higgs doublet model \cite{Eilam:1990zc}, supersymmetry \cite{Yang:1997dk}, technicolor \cite{Lu:2003yr} predict branching ratios for the top quark FCNC decays of the order of $10^{-6}-10^{-5}$. Recent results from CMS experiment place upper bound on the top quark FCNC branching ratio from different channels as $BR(t\to u \gamma)<1.61\times10^{-4}$ and  $BR(t\to c \gamma)<1.82\times10^{-3}$ at $95\%$ confidence level  \cite{Khachatryan:2015att}.

One of the future collider projects currently under consideration after the LHC era is the Future Circular Collider (FCC) \cite{fcc} which includes an option for hadron-electron (FCC-he) collider. This mode is considered to be realized by accelerating electrons up to 60 GeV and colliding them with a beam of protons at the energy of 50 TeV. Recently, search capability and new physics potential of FCC-he collider has been presented in Ref.\cite{Kumar:2015kca}. The $ep$ colliders has a broad top physics potential which can be consulted through Refs.\cite{Cakir:2009rq,Yue:2012kh,Cakir:2013cfx,Bouzas:2013jha,Xiao-Peng:2013nha,Sarmiento-Alvarado:2014eha,Dutta:2013mva,Bouzas:2015rgw,Zhang:2015ado,Liu:2015kkp,Boroun:2015fwa}. Our study is based on FCC-he which would provide sufficient energy to search for top quark FCNC interactions in a clean environment with suppressed backgrounds from strong interaction process \cite{Ohmi:2015njv, Oide:2016mkm}.

In this work, we investigate the anomalous FCNC $tq\gamma$ couplings via single top quark production for probing the FCNC couplings at FCC-he collider. In our study, hadronic decay channel of $W$ boson in the final state of the processes $e^{-}p\to e^{-}W^{\pm}q+X$ and $e^{-}p\to e^{-}W^{\pm}bq+X$ (where $q$ denotes quarks other than top quark) is selected for the signal and background analysis. The event selection and cuts on kinematic variables are discussed in detail. Finally, the discovery potential of anomalous FCNC $tq\gamma$ couplings is examined as a function of luminosity at FCC-he.

\section{Anomalous FCNC interactions}
 The higher order effective operators can be used to describe the BSM effects in model independent way \cite{Grzadkowski:2010es}. For the FCNC $tq\gamma$ couplings the effective Lagrangian can be written as \cite{AguilarSaavedra:2008zc}
\begin{eqnarray}\label{eq1}
 L_{FCNC}&=&
 \frac{g_{e}}{2m_{t}}\bar{u}\sigma^{\mu\nu}(\lambda_{ut}^{L}P_{L}+\lambda_{ut}^{R}P_{R})tA_{\mu\nu} + \frac{g_{e}}{2m_{t}}\bar{c}\sigma^{\mu\nu}(\lambda_{ct}^{L}P_{L}+\lambda_{ct}^{R}P_{R})tA_{\mu\nu}+h.c.
\end{eqnarray}
where  $g_{e}$ is the electromagnetic coupling constant;  $\lambda_{qt}^{L(R)}$ are the strength of anomalous FCNC couplings for $tq\gamma$, which vanish at the lowest order in SM; $P_{L(R)}$ denotes the left (right) handed projection operators; $\sigma^{\mu\nu}$  is the tensor defined as $\sigma^{\mu\nu}=\frac{i}{2}[\gamma^{\mu},\gamma^{\nu}]$ for the FCNC interactions. Here, no specific chirality is assumed for the FCNC interaction vertices, i.e. $\lambda_{q}^{L}=\lambda_{q}^{R}=\lambda_{q}$. 

The effective Lagrangian can be used to calculate both production cross sections and the branching ratios of the $t\to q\gamma$ decays. At present, the observed bounds on the top quark FCNC decays are still rather weak. However, the low energy flavor transitions mediated by top quark loops may also be affected and could therefore provide helpful information for direct searches at high-energy colliders. The top quark FCNC interactions affect $b$ quark FCNC decays through loop diagrams as mentioned in Ref. \cite{Yuan:2010vk,Li:2011fza}. The bounds \cite{Yang:2014efd} on the real FCNC couplings are lower than the current direct limits but still accessible at the high-luminosity run of LHC. In our calculations, we use the effective interaction vertices at the leading order level, however we change its parameters ($\lambda_q$) in an accessible range (0-0.05). More vertices with FCNC couplings each having an order of $\lambda_q= 10^{-2}$ contributes less. 
\section{Production Cross Sections}
The existence of the anomalous $tq\gamma$ couplings can lead to the production of a single top quark in $ep$ collisions. The top quark single production processes are sensitive to the top FCNC interactions in the high energy collisions.  In this section, to make an estimation for the signal, first we calculate cross section for on-shell single top quark production. The signal cross section for the processes  $e^{-}p\to(e^{-}t+e^{-}\bar{t})X$ is given as $3.238\times 10^{-2}$ pb  while for the process $e^{-}p\to(e^{-}t\bar q+e^{-}\bar{t}q)X$ the cross section is $8.106\times10^{-3}$ pb for equal coupling scenario $\lambda_{u}=\lambda_{c}=0.01$ at the center of mass energy $\sqrt{s_{ep}}\simeq3.46$ TeV of the FCC-he collider. The signal cross sections are given in Table~\ref{tab1}  and Table~\ref{tab11} for the couplings $\lambda_{u}$ and $\lambda_{c}$ in the range of $(0-0.01)$. 
For the cross section calculations, we use \verb|MadGraph5_aMC@NLO| \cite{Alwall:2014hca} in which the effective FCNC couplings is implemented through FeynRules package \cite{Degrande:2011ua} via the Lagrangian described in Eq.~\ref{eq1}. We have used the parton distribution function NNPDF23 \cite{Ball:2012cx} which is already available within the MadGraph 5. In the calculation we used fixed renormalization and factorization scales at $m_Z$ for the pdf used both in \verb|MadGraph5_aMC@NLO| and Pythia 6 \cite{Sjostrand:2006za}. We obtain the cross section $\sigma_{c}=7.58$ fb ($\sigma_{u}=24.88$ fb) for the process $e^{-}p\to(e^{-}t+e^{-}\bar{t})X$, and $\sigma_{c}=2.96$ fb ($\sigma_{u}=5.15$ fb) for the process  $e^{-}p\to(e^{-}t\bar q+e^{-}\bar{t}q)X$ for couplings $\lambda_{u}=0$ and $\lambda_{c}=0.01$ ($\lambda_{c}=0$ and $\lambda_{u}=0.01$), respectively. The cross section depends on $\lambda_u$ and $\lambda_c$ with different strength due to proton parton distribution function.
\begin{figure}
\includegraphics[scale=0.3]{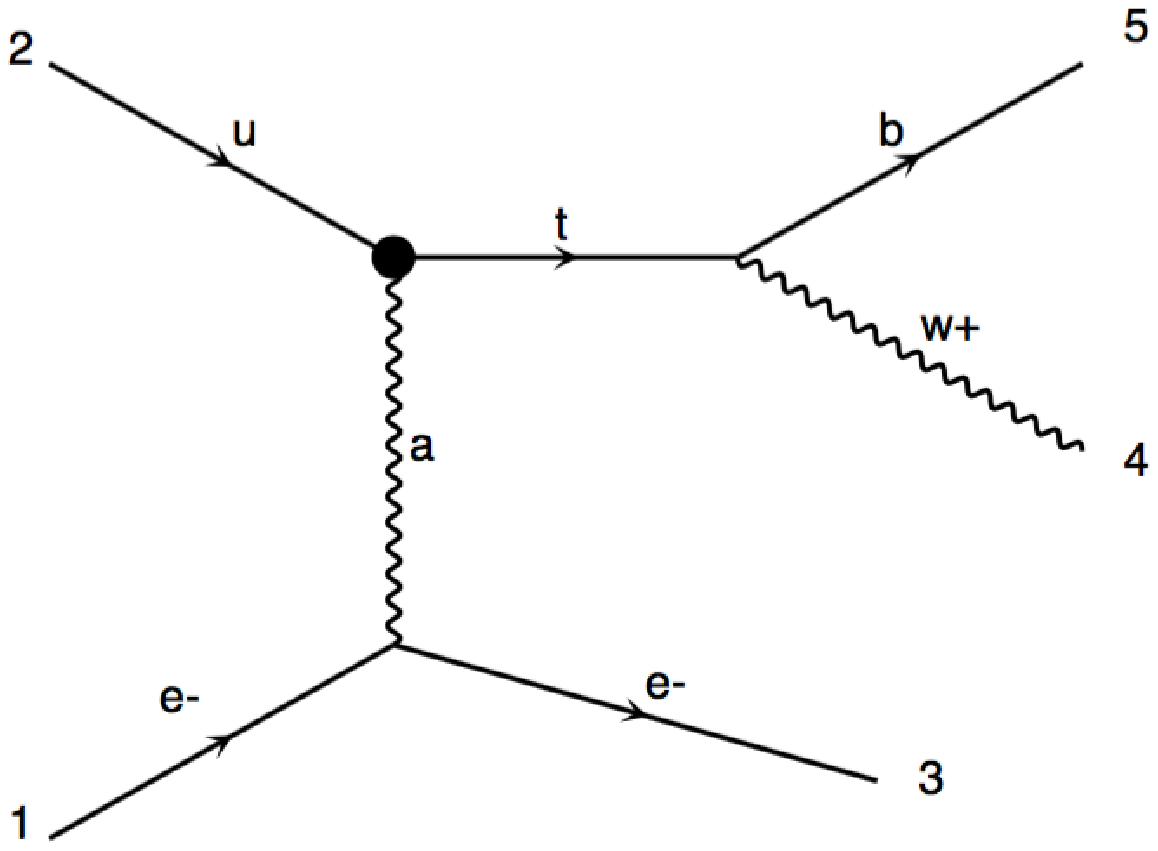}\hspace{0.6cm}\includegraphics[scale=0.3]{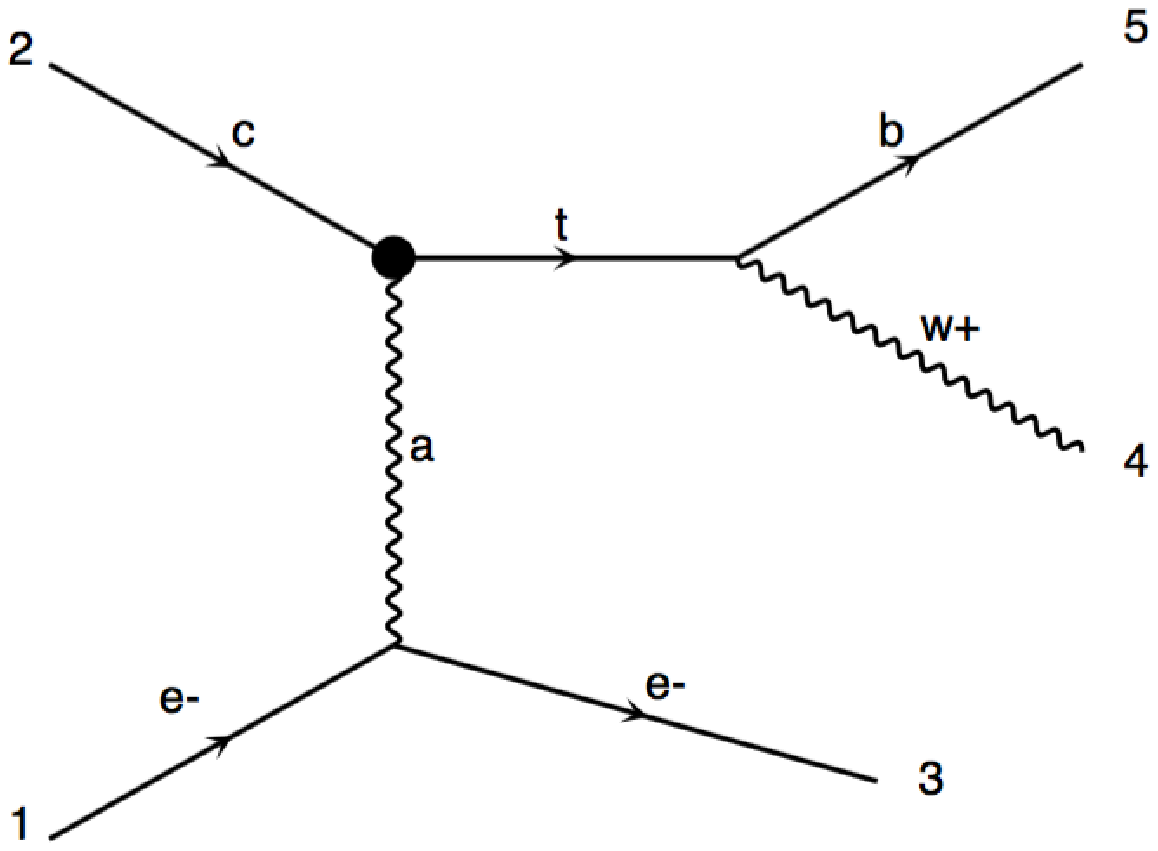}\\
\includegraphics[scale=0.2]{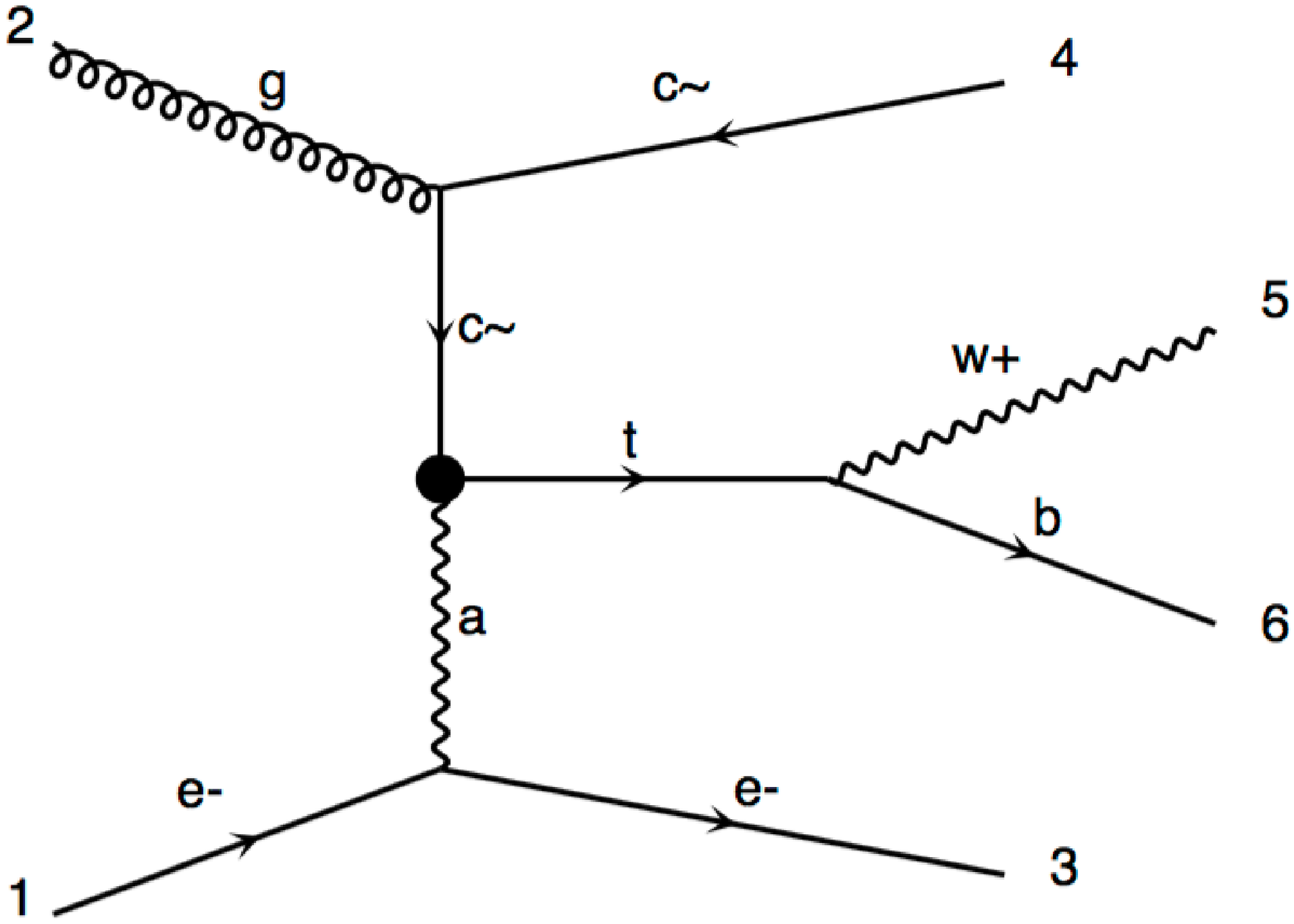}\includegraphics[scale=0.2]{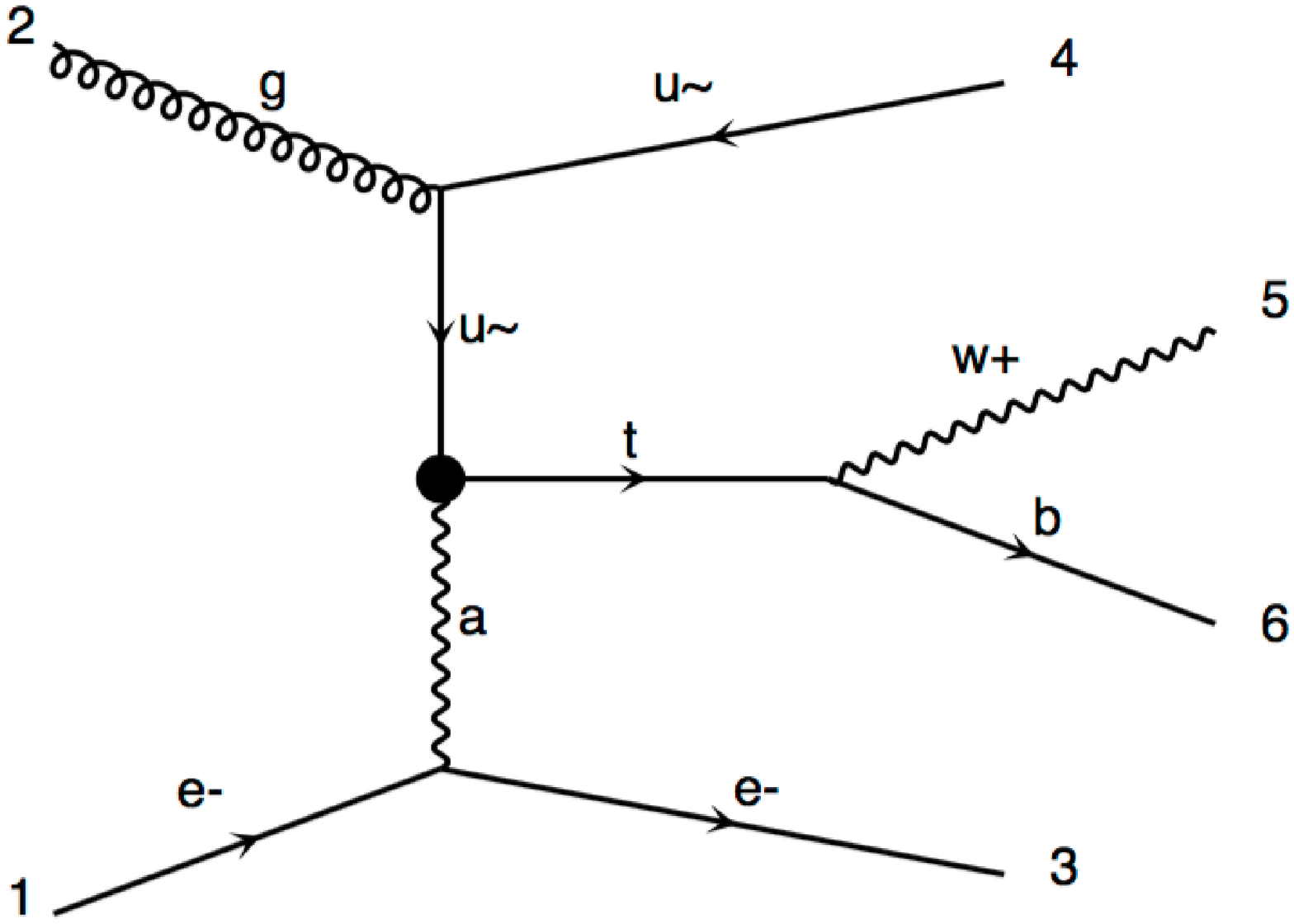}\includegraphics[scale=0.2]{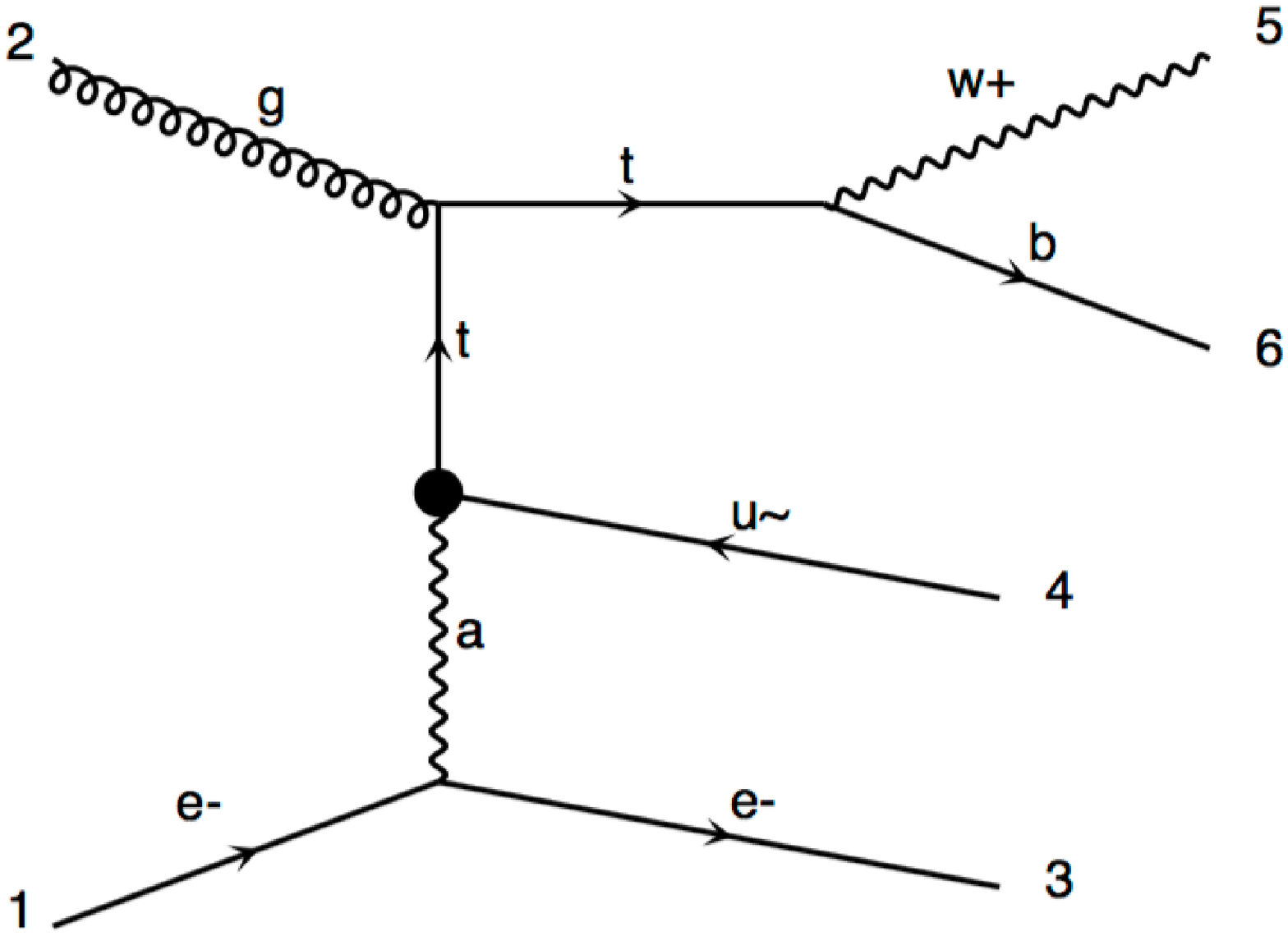}\includegraphics[scale=0.2]{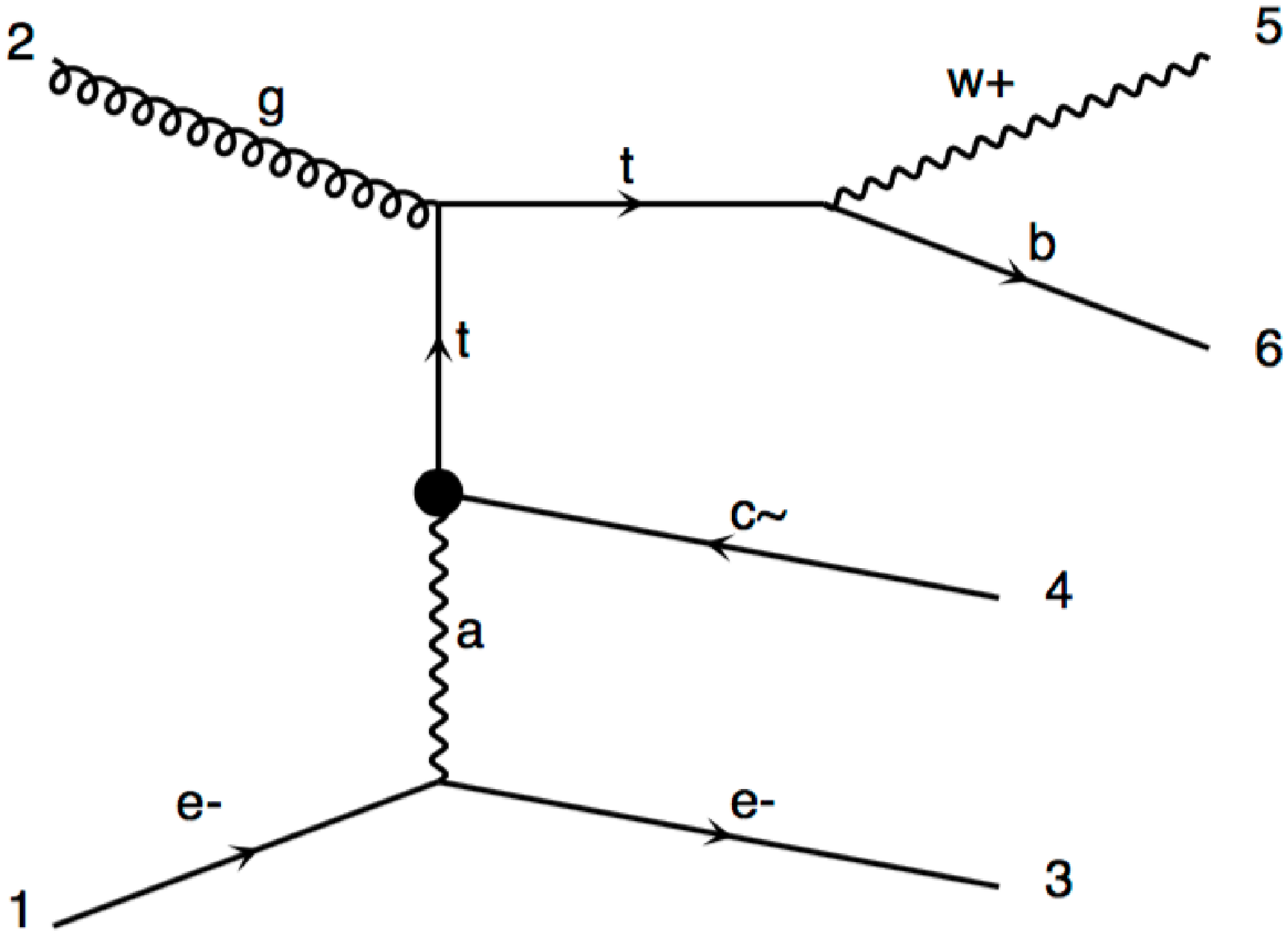}
\caption{Feynman diagrams for single top quark production through FCNC vertices and the top quark decays via charged current. First two diagrams correspond to subprocess $e^- q \to e^- W^+ b$ while others correspond to $e^- g \to e^- q W^+ b$ which contributes to the signal process. 
 \label{fig2}}
\end{figure}

\begin{table}
\caption{The signal cross section values (in pb) for the process $e^{-}p\to(e^{-}t+e^{-}\bar{t})X$
at FCC-he. \label{tab1}}
\begin{ruledtabular}
\begin{tabular}{lccc}
FCC-he  & $\lambda_{c}=10^{-2}$ &$\lambda_{c}=10^{-3}$&$\lambda_{c}=0$ \\\hline 
$\lambda_{u}=10^{-2}$& $3.238\times10^{-2}$  & $2.490\times10^{-2}$ &2.488$\times10^{-2}$ \tabularnewline
$\lambda_{u}=10^{-3}$ & $7.834\times10^{-3}$& $3.243\times10^{-4}$& 2.480$\times10^{-4}$\tabularnewline
$\lambda_{u}=0$  &$7.576\times10^{-3}$ & $7.580\times10^{-5}$ &  0\tabularnewline
\end{tabular}
\end{ruledtabular}
\end{table}

\begin{table}
\caption{The signal cross section values (in pb) for the process $e^{-}p\to(e^{-}t\bar q+e^{-}\bar{t}q)X$
at FCC-he. \label{tab11}}
\begin{ruledtabular}
\begin{tabular}{lccc}
FCC-he  & $\lambda_{c}=10^{-2}$ &$\lambda_{c}=10^{-3}$&$\lambda_{c}=0$ \\\hline 
$\lambda_{u}=10^{-2}$& $8.106\times10^{-3}$  & $5.161\times10^{-3}$ &5.150$\times10^{-3}$ \tabularnewline
$\lambda_{u}=10^{-3}$ & $3.032\times10^{-3}$& $8.132\times10^{-5}$& 5.142$\times10^{-5}$\tabularnewline
$\lambda_{u}=0$  &$2.957\times10^{-3}$ & $2.973\times10^{-5}$ &  0\tabularnewline
\end{tabular}
\end{ruledtabular}
\end{table}

\section{Signal and Background Analysis}
\begin{figure}
\includegraphics[scale=0.5]{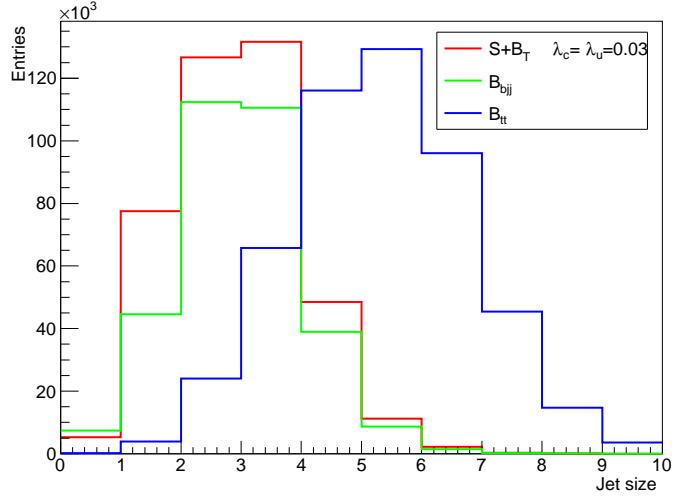}
\caption{The distribution of the jet size in signal events, and also in the important backgrounds; $B_{tt}$: $e^-p\to e^- t \bar t+X$, $B_{bjj}$: $e^-q_b\rightarrow e^- q_b j_2+X$, with $q_b=b$ or 
 $\bar{b}$ and $j_2=q\bar{q}$ or $gg$.\label{fig21}}
\end{figure}

In this section, the analysis of FCNC $tq\gamma$ couplings through the signal processes $e^-p\to e^- W^{\pm} q+X$ and $e^{-}p\to e^{-}W^{\pm}bq+X$ as well as relevant backgrounds at FCC-he are given. While the first process includes both the signal and the interfering background, the second process includes only signal. In the analysis, we take into account off-shell top quark FCNC interaction vertices ($tq\gamma$). The Feynman diagrams for the signal processes are shown in Fig.~\ref{fig2}. The signal processes are studied through the on-shell $W$ boson production and $W$ boson decays hadronically, the characterization of the signal processes are given by the presence of at least three jets and an electron in the final state. In order to generate signal and background events we use \verb|MadGraph5_aMC@NLO| \cite{Alwall:2014hca}. For the signal the effective Lagrangian described by Eq.~\ref{eq1} with FCNC couplings is implemented through FeynRules package \cite{Alloul:2013bka} into the  \verb|MadGraph5_aMC@NLO| as a Universal FeynRules Output (UFO) module \cite{Degrande:2011ua}. Pythia 6 and Delphes 3 \cite{deFavereau:2013fsa} are used for parton showering, hadronization and fast detector simulation, respectively. Jets are clustered using FastJet \cite{Cacciari:2011ma} with the anti-kt algorithm \cite{Cacciari:2008gp} where a cone radius is used as $R=0.5$. In our analysis, b-tagging with efficiency 75\% plays an important role to select final state. Misidentification probability of light quark and c quark as b-jet is taken to be 0.1\% and 5\%, respectively. In order to distinguish signal and background, we apply the kinematic selection cuts as shown in Table~\ref{tab2}. At least three jets are required and an electron is selected in the event with transverse momentum $p_T>20$ GeV. The  distribution of the number of jets in signal events for $\lambda_q=0.03$, and also in the most important backgrounds is given in Fig.~\ref{fig21}.
One of the three jets is tagged as the b-jet while the others are used to reconstruct W boson-mass. The b-tagged jet with $p_T>40$ GeV and other two jets with $p_T>$30 GeV are considered. 
\begin{table}
\caption{Kinematic cuts used for the analysis of signal and background events. Pre-selection cuts are used to select the events with three jets and one electron with transverse momentum grater than 20 GeV. \label{tab2}}
\begin{ruledtabular}
\begin{tabular}{ccc}
Cuts & Definitions \\ \hline 
Cut-0 &  pre-selection cuts with number of jets $\geqslant3$ and one electron with $p^e_T>20$ GeV &\\
Cut-1 & one jet with $b$-tagging& \\
Cut-2 & $p_{T}^b > 40$ GeV and $p_{T}^{j_{2},\,j_{3}} > 30$ GeV, &\\
Cut-3 & $-5 < \eta^{b,\,j_{2},\,j_{3}}<0$ and  $-2.5<\eta^{e} < 2.5$&\\
Cut-4 & 60 GeV $< M_{inv}^{rec}(j_{2},j_{3})< 90$ GeV& \\
Cut-5 &130 GeV $< M_{inv}^{rec}(j_{b},j_{2},j_{3})< 200$ GeV& \\
\end{tabular}
\end{ruledtabular}
\end{table}

Due to the energy asymmetry of the collider pseudo-rapidity of the jets mainly peaked backward (or forward) region depending on $ep$ (or $pe$) collisions, therefore it is taken to be in the interval $-5<\eta<0$ for jets and $-2.5<\eta<2.5$  for the electron. To reconstruct W boson from other two jets, invariant mass of them is required to be between 60 GeV and 90 GeV. As a final cut reconstructed top quark mass from a b-jet and two other jets is selected to be in the range 130 GeV$-$ 200 GeV to count events for further analysis to evaluate the significance for FCNC couplings. After the applied cuts already defined in Table~\ref{tab2}, the number of signal and all relevant backgrounds are given in Table~\ref{tab3}. In Table~\ref{tab3}, $S+B_W$ is defined as the signal for both processes and interference background in $e^-p\to e^- W^{\pm} q+X$.  Since our signal processes include on-shell $W$-boson and its decay into two jets, we classified the background according to $e+V+jets$ which include $eWj$, $eZj$ and we also consider the $eHj$, $ebjj$ and $et\bar t$ backgrounds.

\begin{table}
\caption{The number of signal and relevant background events after each kinematic cuts in the analysis with $L_{int}=$100 fb$^{-1}$.\label{tab3} }
\begin{ruledtabular}
\begin{tabular}{ccccccc}
Processes& Cut-0&Cut-1&Cut-2&Cut-3&Cut-4&Cut-5 \\
\hline 
$S+B_W$($\lambda=0.03$)& 206373& 11687 & 8665 & 7964  &2867 &1883  \\
$S+B_W$($\lambda=0.01$)& 200135 & 7827 & 5776&  5312 &1396 &622  \\
$B_W$&199678 & 7411  & 5447 & 4990& 1184&443  \\
$B_H$&  2279& 979& 802 & 757 &107 &47  \\
$B_Z$& 13420& 1639  & 1145 & 956 &246&110  \\
$B_{tt}$& 9752 & 5594 &  5339 &   4974 &1079 & 460  \\
$B_{bjj}$& 48241 &  17287 &   9936 &   9074 &2573 & 1170  \\
\end{tabular}
\end{ruledtabular}
\end{table}
The relevant backgrounds are defined as $B_W$ for process $e^-p\to e^- W^{\pm} q+X$, $B_Z$ for $e^-p\to e^- Z q+X$, $B_H$ for $e^-p\to e^- H q+X$, $B_{tt}$ for $e^-p\to e^- t \bar t+X$, $B_{bjj}$ for $e^-
 q_b\rightarrow e^- q_b j_2+X$ with $q_b=b$ or 
 $\bar{b}$ and $j_2=q\bar{q}$ or $gg$. The irreducible SM background $B_{bjj}$ is related to $2\to4$ process which includes both off-shell $W$ and $Z$ background as well as $e+3jets$ backgrounds. Total background will be $B_T\equiv$$B_{tt}$+$B_W$+$B_Z$+$B_H$+$B_{bjj}$.  The number of events for relevant backgrounds after Cut-5 are found to be 1170, 460, 443, 110, and  47 for $B_{bjj}$, $B_{tt}$, $B_W$, $B_Z$, and $B_H$ respectively, for the integrated luminosity $L_{int}=100 fb^{-1}$. For the signal and background ($B_W$) we obtain 622 events after Cut-5 for FCNC coupling $\lambda_q=0.01$. The major contribution to the background comes from $B_{bjj}$, even only one b-tag is required in the final state. The number of background events  relatively depend on the branching into the jets.

\begin{figure}
\includegraphics[scale=0.4]{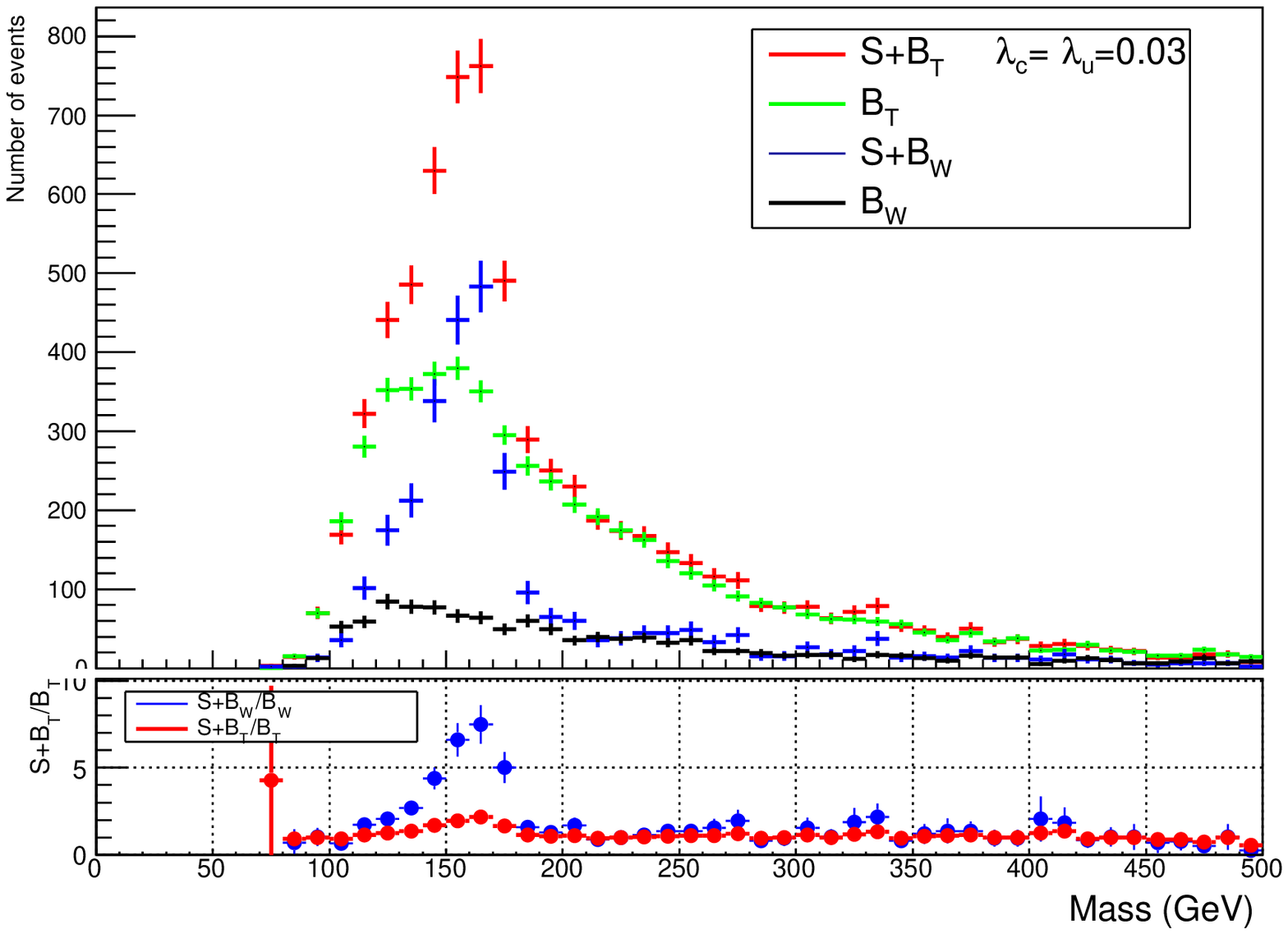}
\includegraphics[scale=0.4]{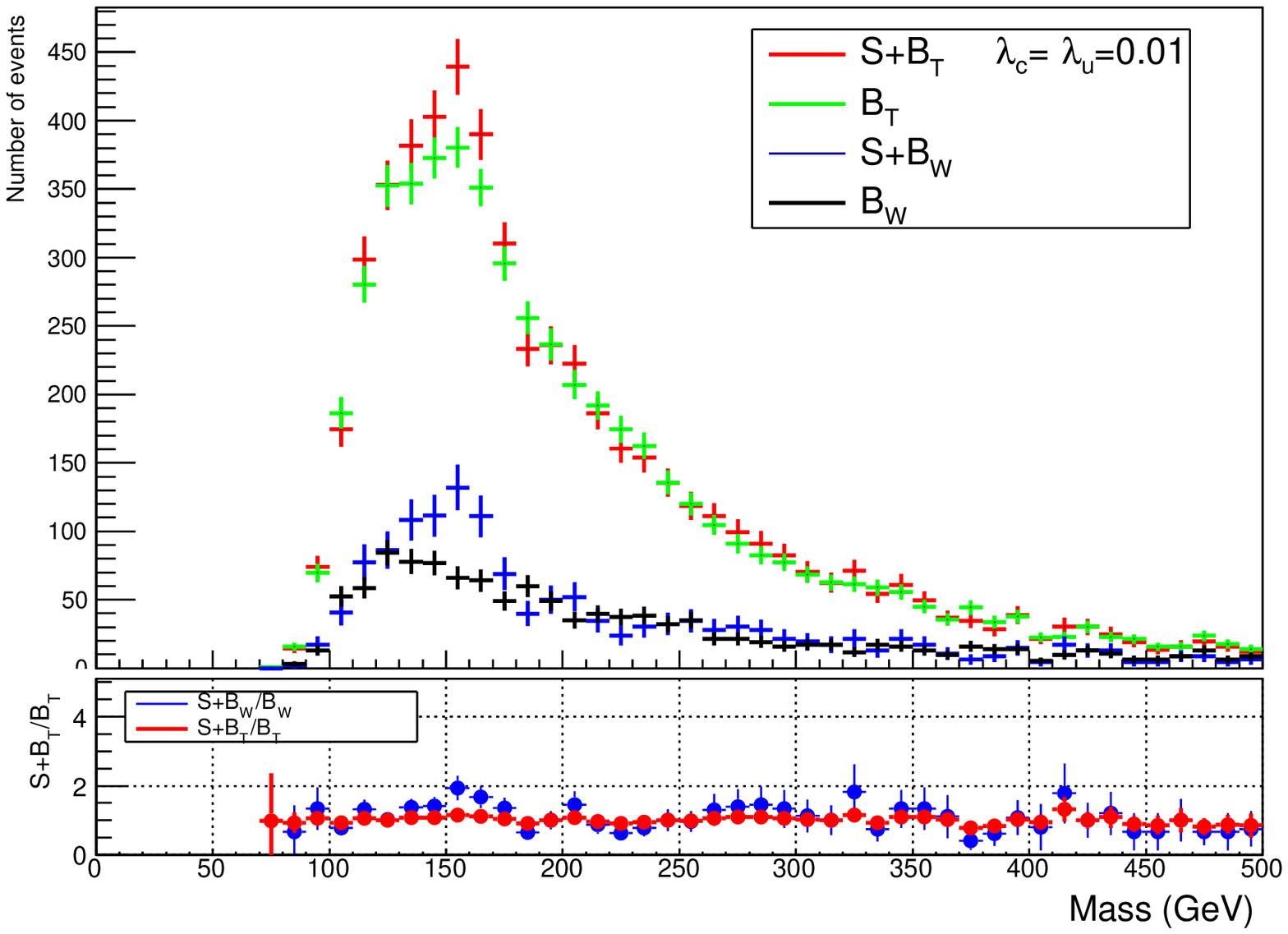}
\caption{Distributions of reconstructed invariant mass of top quark plots for signal and relevant backgrounds with different anomalous FCNC couplings. The lower part of each plot shows the relative ratio of $(S+B)$ and $B$.  \label{fig3}}
\end{figure}
Fig.~\ref{fig3} shows the distribution of reconstructed invariant mass of top quark after Cut-4 for different FCNC couplings when both $\lambda_u$ and $\lambda_c$ are equal. The left plot shows when both $\lambda$ equal to 0.03 for signal and all relevant background are plotted as well as the ratio $(S+B_W)/B_W$ at the bottom of each one. As it can be seen from ratio plots in Fig.~\ref{fig3}, even for a small coupling signal is promoted nearly above the total background. According to the inclusion of all relevant backgrounds ($B_T$) the ratio ($(S+B_T)/B_T$) at the top quark mass decreases a factor about 0.27 for $\lambda$=0.03 when compared with the respective ratio for $B_W$. 

\begin{figure}
\includegraphics[scale=0.6]{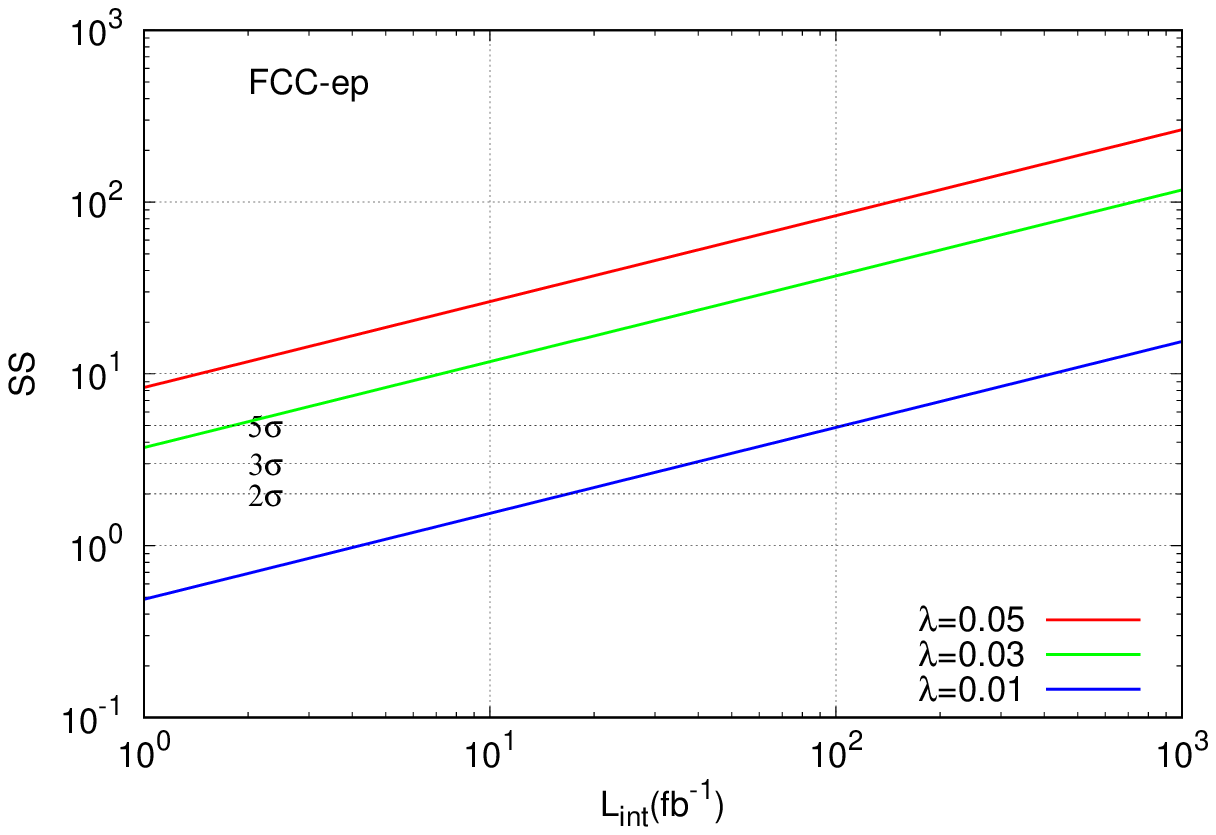}\includegraphics[scale=0.6]{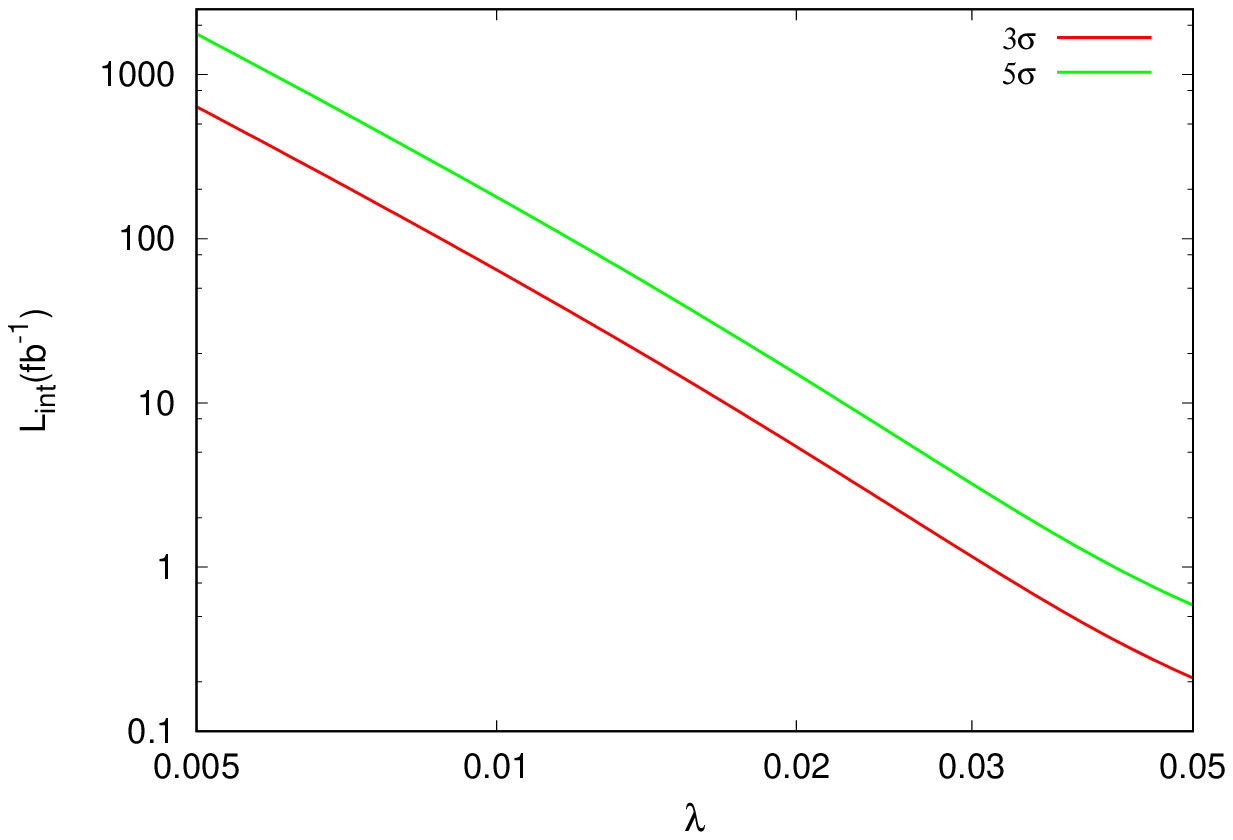}
\caption{On the left the statistical significance depending on integrated luminosity for different anomalous FCNC couplings ($\lambda$). On the right integrated luminosity versus anomalous FCNC couplings at $3\sigma$ and $5\sigma$ significance. \label{fig4}}
\end{figure}

The Statistical Significance ($SS$) is calculated after final cut by using Poisson formula
\begin{equation}
SS=\sqrt{2[(S+B_T)\ln(1+S/B_T)-S]}
\end{equation}
where $S$ and $B_T$ are the signal and total background events at a particular luminosity. Since the proton beam energy is very large, sensitivity to the $\lambda_u$ and $\lambda_c$ couplings are close to each other. The results for the $SS$ values depending on the integrated luminosity (on the left) for equal coupling scenario are given in Fig.~\ref{fig4}. The integrated luminosity versus FCNC couplings (on the right) at $3\sigma$ and $5\sigma$ significance is presented in Fig.~\ref{fig4}. It is clear from Fig.~\ref{fig4} that even at a luminosity of 20 fb$^{-1}$ the FCC-he would provide $2\sigma$
significance for $\lambda_q$=0.01, while for an integrated luminosity of 100 fb$^{-1}$ we obtain $5\sigma$ significance at this coupling. With all the relevant backgrounds, we find $3\sigma$ signal significance results to reach an upper limit $\lambda=0.01$ at the FCC-he with an integrated luminosity of 40 fb$^{-1}$. One can reach at a lower limit of $\lambda=0.005$  for an observability at the integrated luminosity projection of 1 $ab^{-1}$ when it is extrapolated as shown in the right panel of the Fig.~\ref{fig4} .

There are alternative use of effective coupling constants appearing in the effective Lagrangian.  We express our results
in terms of branching ratios which can be comparable with the results
of other studies. Using top quark FCNC decay widths and total decay width we can calculate the branching ratio $BR(t\to q\gamma)$ depending on coupling $\lambda_q$. 
In order to translate the bounds, the branching
ratio is defined as 
\begin{equation}
BR(t\to q\gamma)=\frac{\Gamma(t\to q\gamma)}{\Gamma(t\to q'W^{+})+\Gamma(t\to u\gamma)+\Gamma(t\to c\gamma)}\label{eq:3}
\end{equation}
In this equation, we indicate the tree-level prediction for the top
quark ($t$) decay width into a massless down sector quark ($q'$)
and a $W$-boson,
\begin{equation}
\Gamma(t\to q'W^{+})=\frac{\alpha_{e}}{16\sin^{2}\theta_{w}}|V_{tq'}|^{2}\frac{m_{t}^{3}}{m_{W}^{2}}\left[1-3\frac{m_{W}^{4}}{m_{t}^{4}}+2\frac{m_{W}^{6}}{m_{t}^{6}}\right]\label{eq:4}
\end{equation}
For the total decay width of the top quark, the main contribution comes from the decay $t\to bW$  with the latest value of about $\Gamma(t\to bW)=1.41$ GeV \cite{Patriagnani}, because the
$V_{tb}$ element of CKM matrix is much larger than $V_{ts}$ and
$V_{td}$. The partial widths for the FCNC decay
channels $t\to q\gamma$ are calculated as $
\Gamma(t\to q\gamma)=(1/8)\alpha_{e}\lambda_{q}^{2}m_{t}$.


 The FCNC coupling $\lambda$=0.01 can be converted to the branching ratio $BR(t\to q\gamma)=2\times10^{-5}$ by using Eqs.~(\ref{eq:3})-(\ref{eq:4}) and the partial widths for the FCNC decay channels. We obtain smaller branching ratio when compared with previous $ep$ experiments H1 \cite{Aaron:2009vv} and ZEUS \cite{Chekanov:2003yt} at HERA where they reported limits on the branchings 0.64\% and 0.29\% at 95 \% C.L., respectively.  At a future $ep$ collider project LHeC \cite{AbelleiraFernandez:2012cc} planned to run concurrently with the HL-LHC, the upper limits on branching ratios are the order of  $10^{-5}$ for an integrated luminosity of 100 fb$^{-1}$ \cite{TurkCakir:2017rvu}. We also compare our results on the branching ratios with the LHC results. Based on proton-proton collisions at 8 TeV within the CMS detector at the LHC at an integrated luminosity of 19.8 fb$^{-1}$, the limits on the top quark FCNC branching ratios are $BR(t\to u\gamma)=1.7\times10^{-4}$ and $BR(t\to c\gamma)=2.2\times10^{-3}$ at 95\% C.L. \cite{Khachatryan:2015att}. Our limit on the branching ratio is one order smaller than the LHC Run-I reach. The projected limits on top FCNC couplings at LHC 14 TeV and HL-LHC have been reported in Ref. \cite{ATLAS:2013hta}, where the expected upper limits on branching ratio $t\to q\gamma$ are $8\times10^{-5}$ and $2.5\times10^{-5}$ for an integrated luminosity 300 fb$^{-1}$ and 3000 fb$^{-1}$, respectively.  

\section{Conclusions}
We conclude that FCC-he, with an electron energy of 60 GeV and a proton energy of 50 TeV, would provide significant single top quark production event rates via investigated channel. Top quark FCNC couplings ($\lambda>0.01$) can be searched at the level of significance greater than $3\sigma$ with an integrated luminosity of larger than 40 fb$^{-1}$ at the projected FCC-he. Since b-tagging has an important role for our study, for a more realistic b-tagging efficiency of 60\%, statistical significance decreases about 10\%, and it has also similar effect on the limits of couplings. With our analysis for 1 ab$^{-1}$ the sensitivity to the branching ratio is better than the available experimental limits, and comparable or even better then their projected upgrade results.

\begin{acknowledgments}
We  acknowledge exciting discussion within the FCC-he/LHeC Top physics group. O.Cakir's work was partially supported by Ankara University Scientific Research Projects under the Project No. 16L0430018.
\end{acknowledgments}


\begin{thebibliography}{99}

\bibitem{Glashow70} S.~L.~Glashow, J.~Iliopoulos, and L.Maiani, Phys. Rev. D \textbf{2}, 1285 (1970).
%
\bibitem{Eilam:1990zc} 
  G.~Eilam, J.~L.~Hewett and A.~Soni,
  Phys.\ Rev.\ D {\bf 44}, 1473 (1991)
  Erratum: [Phys.\ Rev.\ D {\bf 59}, 039901 (1999)].

\bibitem{Yang:1997dk} 
  J.~M.~Yang, B.~L.~Young and X.~Zhang,
  Phys.\ Rev.\ D {\bf 58}, 055001 (1998) [hep-ph/9705341].
\bibitem{Lu:2003yr} 
  G.~r.~Lu, F.~r.~Yin, X.~l.~Wang and L.~d.~Wan,
  Phys.\ Rev.\ D {\bf 68}, 015002 (2003)
  [hep-ph/0303122].
 
\bibitem{Khachatryan:2015att} 
  V.~Khachatryan {\it et al.} [CMS Collaboration],
  JHEP {\bf 1604}, 035 (2016)
  [arXiv:1511.03951 [hep-ex]].
\bibitem{fcc}
More information is available on the FCC Web site: https://fcc.web.cern.ch
\bibitem{Kumar:2015kca} 
  M.~Kumar, X.~Ruan, R.~Islam, A.~S.~Cornell, M.~Klein, U.~Klein and B.~Mellado,
  Phys.\ Lett.\ B {\bf 764}, 247 (2017)
  [arXiv:1509.04016 [hep-ph]].
  
 \bibitem{Cakir:2009rq} 
  I.~T.~Cakir, O.~Cakir and S.~Sultansoy,
  Phys.\ Lett.\ B {\bf 685}, 170 (2010)
  [arXiv:0911.4194 [hep-ph]].

\bibitem{Yue:2012kh} 
  C.~X.~Yue, J.~Guo, J.~Zhang and Q.~G.~Zeng,
  Commun.\ Theor.\ Phys.\  {\bf 58}, 711 (2012)
  [arXiv:1203.3627 [hep-ph]].

\bibitem{Cakir:2013cfx} 
  I.~T.~Cakir, A.~Senol and A.~T.~Tasci,
  Mod.\ Phys.\ Lett.\ A {\bf 29}, 1450021 (2014)
  [arXiv:1301.2617 [hep-ph]].

\bibitem{Bouzas:2013jha} 
  A.~O.~Bouzas and F.~Larios,
  Phys.\ Rev.\ D {\bf 88}, no. 9, 094007 (2013)
[arXiv:1308.5634 [hep-ph]].

\bibitem{Xiao-Peng:2013nha} 
  L.~Xiao-Peng, G.~Lei, M.~Wen-Gan, Z.~Ren-You, H.~Liang and S.~Mao,
  Phys.\ Rev.\ D {\bf 88}, no. 1, 014023 (2013)
  [arXiv:1307.2308 [hep-ph]].

\bibitem{Sarmiento-Alvarado:2014eha} 
  I.~A.~Sarmiento-Alvarado, A.~O.~Bouzas and F.~Larios,
  J.\ Phys.\ G {\bf 42}, no. 8, 085001 (2015)
  [arXiv:1412.6679 [hep-ph]].

\bibitem{Dutta:2013mva} 
  S.~Dutta, A.~Goyal, M.~Kumar and B.~Mellado,
  Eur.\ Phys.\ J.\ C {\bf 75}, no. 12, 577 (2015)
  [arXiv:1307.1688 [hep-ph]].

\bibitem{Bouzas:2015rgw} 
  A.~O.~Bouzas and F.~Larios,
  J.\ Phys.\ Conf.\ Ser.\  {\bf 651}, no. 1, 012004 (2015).
  
\bibitem{Zhang:2015ado} 
  Z.~Zhang [LHeC Study Group],
  PoS EPS {\bf -HEP2015}, 342 (2015)
  [arXiv:1511.05399 [hep-ex]].

\bibitem{Liu:2015kkp} 
  W.~Liu, H.~Sun, X.~Wang and X.~Luo,
  Phys.\ Rev.\ D {\bf 92}, no. 7, 074015 (2015)
  [arXiv:1507.03264 [hep-ph]].
  
  \bibitem{Boroun:2015fwa} 
  G.~R.~Boroun,
  Chin.\ Phys.\ C {\bf 41}, 013104 (2017)
  [arXiv:1510.02914 [hep-ph]].

\bibitem{Ohmi:2015njv} 
  K.~Ohmi and F.~Zimmermann, Proceedings, 6th International Particle Accelerator Conference (IPAC 2015) : Richmond, Virginia, USA, May 3-8, 2015.
\bibitem{Oide:2016mkm} 
  K.~Oide {\it et al.},
  Phys.\ Rev.\ Accel.\ Beams {\bf 19}, no. 11, 111005 (2016)
  [arXiv:1610.07170 [physics.acc-ph]].
\bibitem{Grzadkowski:2010es} 
  B.~Grzadkowski, M.~Iskrzynski, M.~Misiak and J.~Rosiek,
  JHEP {\bf 1010}, 085 (2010)
  [arXiv:1008.4884 [hep-ph]].

\bibitem{AguilarSaavedra:2008zc} 
  J.~A.~Aguilar-Saavedra,
  Nucl.\ Phys.\ B {\bf 812}, 181 (2009)
  [arXiv:0811.3842 [hep-ph]].
  
  \bibitem{Yuan:2010vk} 
  X.~Yuan, Y.~Hao and Y.~Yang,
  Phys.\ Rev.\ D {\bf 83}, 013004 (2011)
  [arXiv:1010.1912 [hep-ph]].

\bibitem{Li:2011fza} 
  X.~Q.~Li, Y.~D.~Yang and X.~B.~Yuan,
  JHEP {\bf 1108}, 075 (2011)
  [arXiv:1105.0364 [hep-ph]].
  %
\bibitem{Yang:2014efd} 
  Y.~D.~Yang and X.~B.~Yuan,
  Chin.\ Sci.\ Bull.\  {\bf 59}, no. 29-30, 3760 (2014).
  \bibitem{Alwall:2014hca}
  J.~Alwall {\it et al.},
  JHEP {\bf 1407} (2014) 079
  [arXiv:1405.0301 [hep-ph]].
  
\bibitem{Degrande:2011ua} 
  C.~Degrande, C.~Duhr, B.~Fuks, D.~Grellscheid, O.~Mattelaer and T.~Reiter,
  Comput.\ Phys.\ Commun.\  {\bf 183}, 1201 (2012)
  [arXiv:1108.2040 [hep-ph]].

\bibitem{Ball:2012cx} 
  R.~D.~Ball {\it et al.},
  Nucl.\ Phys.\ B {\bf 867}, 244 (2013)
  [arXiv:1207.1303 [hep-ph]].


%
\bibitem{Alloul:2013bka} 
  A.~Alloul, N.~D.~Christensen, C.~Degrande, C.~Duhr and B.~Fuks,
  Comput.\ Phys.\ Commun.\  {\bf 185}, 2250 (2014)
 [arXiv:1310.1921 [hep-ph]].

\bibitem{Sjostrand:2006za} 
  T.~Sjostrand, S.~Mrenna and P.~Z.~Skands,
  JHEP {\bf 0605}, 026 (2006)
  [hep-ph/0603175].

  \bibitem{deFavereau:2013fsa} 
  J.~de Favereau {\it et al.} [DELPHES 3 Collaboration],
  JHEP {\bf 1402}, 057 (2014)
  [arXiv:1307.6346 [hep-ex]].
  
\bibitem{Cacciari:2011ma} 
  M.~Cacciari, G.~P.~Salam and G.~Soyez,
  Eur.\ Phys.\ J.\ C {\bf 72}, 1896 (2012)
  [arXiv:1111.6097 [hep-ph]].

\bibitem{Cacciari:2008gp} 
  M.~Cacciari, G.~P.~Salam and G.~Soyez,
  JHEP {\bf 0804}, 063 (2008)
  [arXiv:0802.1189 [hep-ph]].

\bibitem{Patriagnani} C. Patriagnani et al., (Particle Data Group),
Chin. Phys. C \textbf{40}, 100001 (2016).

\bibitem{Aaron:2009vv} 
  F.~D.~Aaron {\it et al.} [H1 Collaboration],
  Phys.\ Lett.\ B {\bf 678}, 450 (2009)
  [arXiv:0904.3876 [hep-ex]].

\bibitem{Chekanov:2003yt} 
  S.~Chekanov {\it et al.} [ZEUS Collaboration],
  Phys.\ Lett.\ B {\bf 559}, 153 (2003)
  [hep-ex/0302010]. 
  
  \bibitem{AbelleiraFernandez:2012cc} 
  J.~L.~Abelleira Fernandez {\it et al.} [LHeC Study Group],
  J.\ Phys.\ G {\bf 39}, 075001 (2012)

  [arXiv:1206.2913 [physics.acc-ph]].
%
\bibitem{TurkCakir:2017rvu} 
  I.~Turk Cakir, A.~Yilmaz, H.~Denizli, A.~Senol, H.~Karadeniz and O.~Cakir,
  arXiv:1705.05419 [hep-ph].
\bibitem{ATLAS:2013hta} 
  [ATLAS Collaboration],
  arXiv:1307.7292 [hep-ex].
\end{thebibliography}
\end{document}